\documentclass[aps,showpacs, tightenlines,superscriptaddress]{revtex4}

\begin{document}

\title{Extending the class of quantum states that do not allow local realistic description}

\author{Wies{\l}aw Laskowski}
\affiliation{Instytut Fizyki Teoretycznej i Astrofizyki
Uniwersytet Gda\'nski, PL-80-952 Gda\'nsk, Poland}
\author{Tomasz Paterek}
\affiliation{Instytut Fizyki Teoretycznej i Astrofizyki
Uniwersytet Gda\'nski, PL-80-952 Gda\'nsk, Poland}
\author{Marek {\. Z}ukowski}
\affiliation{Instytut Fizyki Teoretycznej i Astrofizyki
Uniwersytet Gda\'nski, PL-80-952 Gda\'nsk, Poland}

\date{\today}

\begin{abstract}
We present the necessary and sufficient condition 
for the violation of a new series of multipartite Bell's inequalities 
with many measurement settings. 
\end{abstract}


\maketitle

Since the trailblazing paper of Greenberger, Horne and Zeilinger \cite{GHZ} we witness an explosion 
of interest in multiqubit correlations. It was shown in \cite{GHZ} that three or more qubit correlations 
can lead to much more drastic invalidation of the concepts of Einstein, Podolski and Rosen \cite{EPR}, 
that is, to a far much stronger version of the Bell theorem than in two qubit correlations.

The original GHZ paper presented a version of Bell's theorem that does not involve inequalities. 
However it was very quickly noticed by Mermin \cite{MERMIN} that Bell inequalities are needed for 
an efficient analysis of the experimental data. Series of multiqubit Bell inequalities involving two 
settings for each party were proposed \cite{MERMIN}. The inequalities reveal exponentially 
growing, with the number of qubits, discrepancy of quantum mechanical predictions with any local realistic, 
or if one likes local variable, theories.

However it was soon noticed, that in the case of four qubits or more, the inequalities involving more than 
two settings for each observer lead to even more drastic discrepancies \cite{ZUK93, ZUK-KASZ97}. Further, 
it was recently shown that two settings per observation site inequalities for correlation functions 
(to be called later ``standard" ones) in the 
case of some pure states are totally inefficient in showing the conflict between local realism and quantum 
mechanics \cite{SCARANI, ZBLW}. What is even more important, this statement is true even 
for the case of the full set of correlation function Bell inequalities for two settings per 
observer \cite{WZuk,WW,ZB}.

Recently new series of multipartite Bell's inequalities were derived \cite{WZ,BLPZ}. 
Parties now can choose between more than two alternative 
dichotomic observables to measure. 
The new inequalities are generalizations of the standard inequalities \cite{WZuk,WW,ZB}. 
It is interesting whether the recent inequalities possess the properties of the standard inequalities \cite{ZB}.
In \cite{ZB} it was shown that one can derive a compact criterion which discriminates whether a given quantum state 
is able to violate any of the standard inequalities.
This criterion involves the correlation tensor.
Can we build a necessary and sufficient condition for violation of the new inequalities \cite{WZ,BLPZ}
in terms of the correlation tensor?

Here we give an answer to this question. 
We derive necessary and sufficient condition for violation of the new inequalities and conclude from it that
the new inequalities give more stringent constraints on local realistic description of quantum predictions 
than the standard ones. 
We also show that there are states which satisfy the standard inequalities and violate the new inequalities. 

We shall present now a method to derive the conditions that must be satisfied by multiqubit density matrices, so 
that the quantum predictions can violate the inequalities belonging to the new, multisetting, family.

\section{$4 \times 4 \times 2$ inequalities}

For the sake of simplicity we shall start with the case where
two parties can choose between four observables, 
$\hat A_1, \hat A_2, \hat  A_3,  \hat A_4$ and $\hat B_1, \hat B_2, \hat B_3, \hat B_4$,
and the third one between two, $\hat C_1, \hat C_2$. 
We denote such a family of inequalities as a $4 \times 4 \times 2$ one. 

Let us denote by $A_1,...,A_4,B_1,...,B_4$ and $C_1,C_2$ the hypothetical local realistic values  that would be 
obtained by the respective local observers if they choose, for a given run of the experiment, one of the allowed observables. 
The values $A$ are for the first observer, $B$ for the second one and $C$ for the last one. The following relations 
were shown in \cite{ZB} to hold for the first two observers
\begin{equation}
A_{12,12;S'} = \sum_{s_1,s_2 = \pm 1} S'(s_1,s_2) (A_1 + s_1A_2)(B_1+s_2B_2) = \pm 4,
\end{equation}
where $S'=\pm 1$ and it is an arbitrary ``sign" function of the indices $s_1,s_2 = \pm 1$.
Similarly one has
\begin{equation}
A_{34,34;S''} = \sum_{s_1,s_2 = \pm 1} S''(s_1,s_2) (A_3 + s_1A_4)(B_3+s_2B_4) = \pm 4.
\end{equation}
Finally it is easy to show that
\begin{equation}
\sum_{s_1,s_2 = \pm 1} S(s_1,s_2) (A_{12,12;S'} + s_1 A_{34,34;S''})(C_1+s_2C_2) = \pm 16, \label{442}
\end{equation}
where again $S(s_1,s_2)$ is a sign function. This is the algebraic identity which generates a family of Bell inequalities of the type derived in \cite{BLPZ}.

If one averages the identity over the runs of the experiment, and introduces the Bell-GHZ correlation functions, 
$E_{klm} = \langle A_k B_l C_m \rangle_{avg}$, the set of the new inequalities emerge.
However, it is easy to notice that if $S(s_1,s_2) = S_1(s_1) S_2(s_2)$, where $S_i$ are again sign functions, 
the identity (\ref{442}) reduces to the one 
which cannot lead to Bell's inequalities which are interesting in the case of the three qubit correlations. 
This is because $\sum_{s_2 = \pm 1} S_2(s_2) (C_1 + s_2 C_2) = \pm 2 C_1$ or $ \pm 2 C_2$, i.e. one has effectively 
no choice of observables for the third party.

There is only one type of sign function $S(s_1,s_2)$ which gives non trivial new Bell inequalities. One must 
put one value of $S(s_1,s_2)$ different than other three values, e.g. $S(+1,+1) = S(+1,-1) = S(-1,+1) = - S(-1,-1)$. 
For other choices of values the sign function is factorable. With the above choice and $S(+1,+1)=1$, 
for the given run of the experiment expression (\ref{442}) gives:
\begin{eqnarray}
&&\sum_{s_1,s_2 = \pm 1} S'(s_1,s_2) (A_1 + s_1A_2)(B_1+s_2B_2)(C_1+C_2)  \nonumber \\
&&+ \sum_{s_1,s_2 = \pm 1} S''(s_1,s_2) (A_3 + s_1A_4)(B_3+s_2B_4)(C_1-C_2)= \pm 8.   \label{nontrivial442}
\end{eqnarray}
After averaging over many runs of the experiment we get a set of Bell inequalities:
\begin{eqnarray}
&& \Big| \sum_{s_1,s_2 = \pm 1} \sum_{k=1,2} \sum_{l=1,2} \sum_{m=1,2} S'(s_1,s_2) s_1^{k-1} s_2^{l-1} E_{klm} \Big| 
 \nonumber \\
&& \Big| + \sum_{s_1,s_2 = \pm 1} \sum_{k=3,4} \sum_{l=3,4} \sum_{m=1,2} S''(s_1,s_2) s_1^{k-1} s_2^{l-1} (-1)^{m-1} E_{klm} 
\Big| \le 8. \label{ineq442}
\end{eqnarray}
There are as many Bell inequalities as different sign functions $S'$ and $S''$. 
The moduli emerge in (\ref{ineq442}) once one realizes that the replacement of $S'$ by $-S'$ and $S''$ by $-S''$ is always possible. 

An interesting observation is that the above set of Bell inequalities is equivalent to a {\em single} 
inequality of the form
\begin{eqnarray}
&& \sum_{s_1,s_2 = \pm 1} \Big| \sum_{k=1,2} \sum_{l=1,2} \sum_{m=1,2} s_1^{k-1} s_2^{l-1} E_{klm} \Big| 
  \nonumber \\
 &&+\sum_{s_1,s_2 = \pm 1} \Big| \sum_{k=3,4} \sum_{l=3,4} \sum_{m=1,2} s_1^{k-1} s_2^{l-1} (-1)^{m-1} E_{klm} 
\Big| \le 8.  \label{one}
\end{eqnarray}
This is because the $S'$ and $S''$ functions can always be such that their sign is opposite to the one of the 
expression in front of them (in other words we have the simple fact that $|a \pm b| \le c$ if and only if $|a|+|b| \le c$). 
This makes the analysis much more simpler. Instead of considering a very vast family of linear inequalities, 
one deals with just one, which is non-linear. 
A similar property has also the full set of standard Bell inequalities \cite{ZB}.

\section{Violation of the $4 \times 4 \times 2$ inequality}

The product $ \langle A_kB_lC_m \rangle_{avg}$ defines the correlation function $E_{klm}$. 
In the case of quantum correlation functions these can be expressed in form of the scalar product of the 
correlation tensor $\hat T$ with the tensor product of the local measurement
settings (represented by a unit vector), i.e. $E_{klm} = \hat T \circ
(\vec n_{A_k} \otimes \vec n_{B_l} \otimes \vec n_{C_m})$. 
Therefore, in the
quantum case, the inequality (\ref{one}) can be re-expressed
in the following way:
\begin{eqnarray}  
\sum_{s_1,s_2=\pm1} \Big| \hat T \circ (\vec n_{A_1} + s_1 \vec n_{A_2}
) \otimes (\vec n_{B_1} + s_2 \vec n_{B_2} ) \otimes (\vec n_{C_1} + \vec
n_{C_2}) \Big|  \nonumber \\
+\sum_{s_1,s_2=\pm1} \Big| \hat T \circ (\vec n_{A_3} + s_1 \vec n_{A_4} ) 
\otimes (\vec n_{B_3}+s_2  \vec
n_{B_4})\otimes (\vec n_{C_1} - \vec n_{C_2}) \Big| \le 8,  \label{qm442} 
\end{eqnarray}
where $\vec n_{X_i}$ denotes parameters the values of which
define local observable $i$ in the laboratory $X$ ($X=\{A,B,C\}$).

Let us put:  
\begin{equation}
\vec n_{A_1} + s_1 \vec n_{A_2} = 2a_{s_1} \vec A(s_1) ,  
\end{equation}
where  \begin{equation}
a_+^2 + a_-^2 = 1,
\end{equation}
and the vectors $\vec A(\pm) $ are of unit norm. We shall generally omit $1$ when using $s=\pm1$ 
as an index numbering specific variables.                            
One can easily show that
\begin{equation}
\vec A(+)\circ \vec A(-) = 0.
\end{equation}
We introduce analogical vectors $\vec B (\pm)$ and $\vec C (\pm)$ for the other pairs of observables. 
Using these new local vectors one  can write (\ref{qm442}) as: 
\begin{eqnarray}
&& \sum_{s_1,s_2=\pm1} | a_{s_1} b_{s_2} c_+  \hat T \circ (\vec
A(s_1)\otimes  \vec B(s_2)\otimes  \vec C(+) )|  \nonumber \\
&& +\sum_{s_1,s_2=\pm1} | a'_{s_1}b'_{s_2}c_- \hat T \circ  ( \vec
A'(s_1)\otimes  \vec B'(s_2) \otimes  \vec C(-)) | \le 1. 
\label{442_within_T_mod}
\end{eqnarray}
Since $c_+^2 + c_-^2 =1$, i.e. $(c_+, c_-)$ is a unit vector, using 
the Cauchy inequality one gets 
\begin{eqnarray}
&& \sum_{s_1,s_2=\pm1} (\hat T \circ (\vec
A(s_1)\otimes  \vec B(s_2)\otimes  \vec C(+) ))^2  \nonumber \\
&& +\sum_{s_1,s_2=\pm1} (\hat T \circ  ( \vec
A'(s_1)\otimes  \vec B'(s_2) \otimes  \vec C(-)))^2 \le 1.
\label{T_mod}
\end{eqnarray}
Therefore, if for any set of local coordinate systems the three particle correlation 
function satisfies above inequality, then it also satisfies (\ref{ineq442}). 
That is, the above condition is sufficient for Bell inequalities (\ref{ineq442}) to hold.
The negation of this condition is necessary for violation of the inequalities (\ref{ineq442}).

Let us introduce a simpler notation. 
Note that $\hat T \circ (\vec
A(s_1)\otimes  \vec B(s_2)\otimes \vec C(+))$ is  a component of the
tensor $\hat T$ with respect to local Cartesian coordinate system with two axes along the vectors $\vec A(\pm)$ 
for the first observer, $\vec B(\pm)$ for the second one, and $\vec C(\pm)$ for the third one. $\hat T \circ (\vec A'(s_1) \vec B'(s_2) \vec C(-))$ 
has the same property. What is very interesting, and contrasts the standard case \cite{ZB}, the components of $\hat T$ can now be taken with respect to a different local 
coordinate systems for the first two observers, namely with axes along $\vec A'(\pm)$ and $\vec B'(\pm)$.
Thus a concise way to write the condition would be that
\begin{equation}
\sum_{k,l=1,2} T_{kl2}^2 + \sum_{k',l'=3,4} T^{2}_{k'l'1} \leq 1. \label{sufficient}
\end{equation}
where primes in the second term denote the fact that the coordinate systems for the first two observers can be chosen 
independently from those in the first term. From the point of view of experiment this means that the plane of 
observations for observables $1$ and $2$ can be different from the plane in which lie the observation direction $3$ and $4$.

Note further, that once (\ref{sufficient}) is satisfied for all possible planes of observations, then the state 
endowed with $\hat T$ is not capable to violate the new inequalities. However, if for given planes of 
observations (\ref{sufficient}) is satisfied, this only means that for such planes one cannot expect violations.

\section{The necessity}

We shall prove that the condition (\ref{sufficient}) is also a necessary one for the new inequalities to hold. 
To this end, we have to prove that the violation of (\ref{sufficient}) leads to the violation of the inequality 
(\ref{ineq442}). 
This can be done by showing that the maximum value of the left hand side of (\ref{ineq442}) is given by the left 
hand side of inequality (\ref{sufficient}).

The three qubit correlation tensor can be Schmidt decomposed into:
\begin{equation}
\hat T = \hat P_1 \otimes \vec \gamma_1 + \hat P_2 \otimes \vec \gamma_2 +
\hat P_3 \otimes \vec \gamma_3, 
\label{schmidt} 
\end{equation}
where the three unit vectors $\vec \gamma_i$ form a basis set in $R^3$ and the (unnormalized) rank two tensors, also 
orthogonal to each other:
\begin{equation}
\hat P_i \circ \hat P_j = 0 \qquad i \ne j.
\end{equation}
We shall assume throughout that the rank two tensors are ordered by their indices
 in accordance with decreasing norms.

Let us define two new tensors
\begin{eqnarray}
&& \hat S' = \sum_{s_1,s_2=\pm1} S'(s_1,s_2) a_{s_1} \vec A(s_1)\otimes b_{s_2}
\vec B(s_2), \nonumber \\  
&& \hat S'' = \sum_{s_1,s_2=\pm1} S''(s_1,s_2)
a'_{s_1} \vec A'(s_1)\otimes b'_{s_2} \vec B'(s_2).  \label{Sdefinition}
\end{eqnarray}
Please note that since $\hat S'$ and $S''$ are Schmidt decomposed itself; it is easy to see that their norm is $1$, 
and is independent of $S'(s_1,s_2)$ and $S''(s_1,s_2)$. Inserting the decomposition of $\hat T$ given by Eq. (\ref{schmidt}) and definitions (\ref{Sdefinition}) 
into the left hand side of ineq. (\ref{442_within_T_mod})
one gets the following inequality:
\begin{eqnarray}
&& \Big|c_+ (\hat P_1 \otimes \vec \gamma_1 + \hat P_2
 \otimes \vec \gamma_2 + 
\hat P_3 \otimes \vec \gamma_3) \circ \hat S' \otimes \vec C(+) \nonumber \\
&& \pm ~ c_- (\hat P_1 \otimes \vec \gamma_1 + \hat P_2 \otimes \vec \gamma_2 + 
\hat P_3 \otimes \vec \gamma_3) \circ \hat S'' \otimes \vec C(-) \Big| \le 1. \label{pre-schm}
\end{eqnarray}
If one specifies the vectors $\vec{C}(\pm)$ so that $\vec \gamma_1 = \vec C(+)$, $\vec \gamma_2 = \vec C(-)$ 
then the value of the left hand side expression is given by
\begin{equation}
c_+ \hat P_1 \circ \hat S' + c_- \hat P_2 \circ \hat S'',
\label{schm}
\end{equation}
and it is easy to see from the properties of Schmidt decomposition that the modulus of this expression is the maximal possible value of (\ref{pre-schm}). One can interpret the expression (\ref{schm}) as the scalar product in a two dimensional vector space. 
The vectors under consideration are: $(c_+, \pm c_-)$ and $(\hat P_1 \circ \hat S',\hat P_2 \circ \hat S'')$.
Since the coefficients $c_{\pm}$ are arbitrary and the norm of the first vector is equal to $1$, 
the maximum value of the expression is given by:
\begin{equation}
\sqrt{(\hat P_1 \circ \hat S')_{max}^2 + (\hat P_2 \circ \hat S'')_{max}^2} \label{max}  
\end{equation}

Since $\hat S'$ and $\hat S''$ are independent, the maximum of this expression is reached when both terms are maximal.  Let us focus on the maximum possible value of $\hat P_1 \circ \hat S'$. The Schmidt decomposition of $P_1$ can 
be written as  
\begin{equation}
\hat P_1 = \sum_{i=1}^3 r_i \vec v_i \otimes \vec w_i.
\end{equation}
All the unit vectors $\vec v_i$ are perpendicular to each other (and so are $\vec w_i$'s), and real numbers $r_i$ are 
ordered as follows: $r_1 \ge r_2 \ge r_3$.
Thus
\begin{equation}
\hat P_1 \circ \hat S' = \sum_{i=1}^3 r_i \vec v_i \otimes \vec w_i ~\circ~ \sum_{s_1,s_2=\pm1} S'(s_1,s_2) a_{s_1} 
\vec A(s_1)\otimes b_{s_2} \vec B(s_2). \label{p1s}
\end{equation}
The following question arises: how can we choose the sign function $S'$ to maximize above expression?
Note that factorizable $S(s_1,s_2)$ can lead to maximum only when $\hat P_1$ is 
factorable itself. Let us put the case aside for a while. Thus we put $S'(s_1,s_2)$ to be non-factorable, e.g. 
$S'(+1,+1)=S'(+1,-1)=S'(-1,+1)=-S'(-1,-1)$. Assume that $S(+1,+1)=1$, in such a case we can write $\hat S'$ as:
\begin{eqnarray*}
&\hat S' =& a_+ \vec A(+) \otimes b_+ \vec B(+) + a_+ \vec A(+) \otimes b_- \vec B(-) \\
&& +~ a_- \vec A(-) \otimes b_+ \vec B(+) - a_- \vec A(-) \otimes b_- \vec B(-) \\
&& =~ a_+ \vec A(+) \otimes \Big\{b_+ \vec B(+) + b_- \vec B(-)\Big\} + a_- \vec A(-) \otimes \Big\{b_+ \vec B(+) - 
b_- \vec B(-)\Big\}.
\end{eqnarray*}
Now expression (\ref{p1s}) reads:
\begin{eqnarray*}
&& \hat P_1 \circ \hat S' = \big(r_1 \vec v_1 \otimes \vec w_1 + r_2 \vec v_2 \otimes \vec w_2 + r_3 \vec
v_3 \otimes \vec w_3 \big) \\  &\circ& \big (a_+ \vec A(+) \otimes \big\{b_+ \vec
B(+) + b_- \vec B(-)\big\} ~+~ a_- \vec A(-) \otimes \big\{b_+ \vec B(+) - b_-
\vec B(-)\big\} \big).  
\end{eqnarray*}
We can utilize the fact that $b_+ \vec B(+) ~\pm~ b_-\vec B(-)$ are unit vectors.
Because the term with the coefficient $r_3$ is the smallest in the Schmidt decomposition of $\hat P_1$, 
and $\hat S'$ has been put to a form in which it has only two terms, thus to maximize the whole 
scalar product we should make two the vectors $b_+ \vec B(+) + b_-\vec B(-)$ and 
$b_+ \vec B(+) - b_-\vec B(-)$
 perpendicular, by e.g. putting $b_+=b_-=\frac{1}{\sqrt{2}}$,
and ignore the smallest term in $\hat P_1$.
Let us denote:
\begin{eqnarray*}  
&& \vec b = \frac{1}{\sqrt{2}} \vec B(+) + \frac{1}{\sqrt{2}} \vec B(-),\\  && \vec
b_{\perp} = \frac{1}{\sqrt{2}} \vec B(+) - \frac{1}{\sqrt{2}} \vec B(-).
\end{eqnarray*}  
Thus we get:
\begin{equation}
 \hat P_1 \circ \hat S' = (r_1 \vec v_1 \otimes \vec w_1 + r_2 \vec v_2 \otimes \vec w_2) \circ (
a_+ \vec A(+) \otimes \vec b + a_- \vec A(-) \otimes
\vec b_{\perp}).  \label{PS}
\end{equation}  
The maximum is attainable for $\vec v_1 = \vec A(+)$,
$\vec v_2 = \vec A(-)$, $\vec w_1 = \vec b$ and $\vec w_2 = \vec b_{\perp}$, 
and $r_1/a_+ = r_2/a_-$.
Since all above vectors are normalized the maximum value of (\ref{PS}) is:
\begin{equation}  
(\hat P_1 \circ \hat S')_{max} = \sqrt{r_1^2 + r_2^2} = \sqrt{\sum_{i,j=1,2} T_{ij2}^2}.  
\end{equation} 
The local coordinates of the correlation tensor are such that for the first observer they are defined by a basis set containing 
$\vec A(+)$ and $\vec A(-)$, for the second observer, the basis set contains $\vec b$ and $\vec b_{\perp}$, 
and for third one $\vec \gamma_1$ and $\vec \gamma_2$.
It is easy to see that the case of factorable $\hat P_1$ can be dealt with by taking the limit of $r_2 \rightarrow 0$.
 
If one writes the Schmidt decomposition of $\hat P_2$ and takes into account that the vectors $\vec A'(\pm)$ and $\vec B'(\pm)$ 
can be chosen totally independently from $\vec A(\pm)$ and $\vec B(\pm)$, then by making analogical steps as for $\hat P_1 \circ \hat S'$ one gets:
\begin{equation}
(\hat P_2 \circ \hat S'')_{max} = \sqrt{\sum_{k',l'=1,2} T_{k'l'1}^2}.  
\end{equation}
After inserting the last two equations into (\ref{max}) one obtains the necessary condition 
for the inequality (\ref{ineq442}) to hold:
\begin{equation}
\sqrt{\sum_{i,j=1,2} T_{ij2}^2 + \sum_{k',l'=1,2}  T_{k'l'1}^2} \le 1.
\end{equation}
One immediately sees that this condition is identical with the sufficient one. 
Thus it is the necessary and sufficient one for violation of the Bell inequality (\ref{ineq442}).

This condition gives more stringent constraints on local realistic description of quantum predictions than 
the sufficient condition for the standard inequalities \cite{ZB} to hold:
\begin{equation}
\sum_{k,l,m=1,2} T^2_{klm} = \sum_{k,l=1,2} T^2_{kl1} + \sum_{k,l=1,2} T^2_{kl2} \le 1. \label{standard_cond}
\end{equation}
The condition (\ref{sufficient}) is more restrictive, because the planes of observations for the first two 
observers do not have to be the same in both terms (in contradistinction with the middle expression 
of ineq.(\ref{standard_cond})).

\section{Examples}

For example consider the generalized GHZ state:
\begin{equation}
| \psi \rangle = \cos\alpha |000 \rangle + \sin\alpha |111 \rangle, \label{genGHZ}
\end{equation}
where $| 0 \rangle$ and $| 1 \rangle$ are the eigenvectors of the $\sigma_z$ operator. 
There is a range $\alpha \in [0,\pi/12]$ within which $|\psi \rangle$ satisfies all standard 
correlation function Bell inequalities 
\cite{SCARANI,ZBLW}, but violates the
multisetting Bell inequalities \cite{WZ,BLPZ}. 
The non vanishing components of the correlation tensor $T_{ijk} = \langle \psi| \sigma_{x_i}^{(1)}
\sigma_{x_j}^{(2)} \sigma_{x_k}^{(3)} | \psi \rangle$ with $x_1\!=\!x$, $x_2\!=\!y$, $x_3\!=\!z$,  for this state are:
$T_{333} \!=\! \cos2\alpha$, $T_{111} \!=\! \sin2\alpha$, $T_{221} \!=\! T_{212} \!=\! T_{122} \!=\! -\sin2\alpha$.
Therefore the condition (\ref{sufficient}) can be put in the following form
\begin{equation}
\sum_{i,j=1,2} T_{ij2}^2 + \sum_{k',l' = 1,3} T_{k'l'3}^{2} = T_{212}^2+T_{122}^2+T_{3'3'3}^{2}.
\end{equation}
By putting the actual values of these parameters one gets:
\begin{equation}
T_{212}^2+T_{122}^2+T_{3'3'3}^{2} = T_{212}^2+T_{122}^2+T_{333}^2 = 1+ \sin^2 2 \alpha > 1. \label{viol442}
\end{equation}
That is, the generalized GHZ state violates the inequality (\ref{ineq442}) for the whole range of $\alpha$,
in contradistinction with the standard case \cite{ZBLW}.

\section{$3 \times 3 \times 2$ case}

In \cite{BLPZ} methods of generation of different type inequalities (than $4 \times 4 \times 2$) were shown.
One can find there the following Bell inequality
\begin{equation}
4|\sum_{m=1,2} E_{11m}| + \sum_{s_1,s_2 = \pm 1} |\sum_{k,l=3,4} \sum_{m=1,2} s_1^{k-1} s_2^{l-1} (-1)^{m-1}E_{klm}| \le 8.
\label{332}
\end{equation}
It involves only three settings for the first two observers.
The condition for quantum correlations to satisfy this inequality reads
\begin{equation}
4 \Big| \hat T \circ \vec n_{A_1} \otimes \vec n_{B_1} \otimes \sum_{k=1,2} \vec n_{C_k} \Big|
+ \sum_{s_1,s_2= \pm 1} 
\Big| \hat T \circ 
\sum_{i=3,4} s_1^{i-1} \vec n_{A_i} \otimes \sum_{j=3,4} s_2^{j-1} \vec n_{B_j} \otimes \sum_{k=1,2} (-1)^{k-1}
\vec n_{C_k} \Big| \le 8.
\label{quant332}
\end{equation}
As earlier, one can introduce new vectors: 
\begin{eqnarray}
2 a_1 \vec A_1 &=& \vec n_{A_3} - \vec n_{A_4}, \\
2 a_2 \vec A_2 &=& \vec n_{A_3} + \vec n_{A_4}, \\
\vec A_3' &=& \vec n_{A_1}.
\end{eqnarray}
The prime over the third vector is put in order to indicate that it does not have to be orthogonal to the other two.
Analogously we define new three vectors on Bob's side, and for the third party:
\begin{eqnarray}
2 c_2 \vec C_2 &=& \vec n_{C_1} - \vec n_{C_2}, \\
2 c_3 \vec C_3 &=& \vec n_{C_1} + \vec n_{C_2}.
\end{eqnarray}
Therefore the components of the correlation tensor with respect to these vectors must satisfy
\begin{equation}
|c_3T_{3'3'3}|+\sum_{k,l=1,2}|a_k b_l c_2 T_{kl2}| \le 1,
\end{equation}
where e.g. $T_{3'3'3} = \hat T \circ \vec A'_3 \otimes \vec B'_3 \otimes \vec C'_3$.
Employing twice the Cauchy inequality one immediately sees that the condition to satisfy (\ref{332}) is
\begin{equation}
T_{3'3'3}^2 + \sum_{k,l=1,2} T_{kl2}^2 \leq 1. \label{332cond}
\end{equation}
Please notice here the very interesting feature: the coefficient $4$ present in 
(\ref{332}) and (\ref{quant332}) has disappeared!
The tensor components enter (\ref{332cond}) with equal weight.

It is easy to show that for the generalized GHZ states the value of the 
expression (\ref{332cond}) can be for any $ 0 < \alpha \leq \frac{\pi}{12}$ equal to $1+\sin^2 2\alpha > 1$.

\section{$4 \times 4 \times ... \times 4 \times 2$ inequality}

The $4 \times 4 \times ... \times 4 \times 2$ inequality for $N$ particle systems introduced in \cite{BLPZ} reads
\begin{eqnarray}
&& \sum_{s_1...s_{N-1} = \pm 1}
\Big| \sum_{i_1,...,i_N = 1,2} s_1^{i_1-1}...s_{N-1}^{i_{N-1}-1}(-1)^{i_N-1} E_{i_1...i_N} \Big|  \nonumber \\
&& +\sum_{s_1...s_{N-1} = \pm 1}
\Big| \sum_{k_1,...,k_{N-1} = 3,4} \sum_{k_N=1,2} s_1^{k_1-1}...s_{N-1}^{k_{N-1}-1} E_{k_1...k_N} \Big|
\le 2^N.
\end{eqnarray}
After the same procedure as for $4 \times 4 \times 2$ (special case of this more general one) one gets the necessary 
condition to satisfy this inequality:
\begin{equation}
\sum_{i_1,....i_{N-1}=1,2} T_{i_1...i_{N-1}1}^2 + 
\sum_{k_1...k_{N-1}=1,3} T_{k_1'...k_{N-1}'2}^{2} \le 1. \label{N_necessary}
\end{equation}

This condition can be violated by all $N$ particle generalized GHZ states:
$\cos \alpha | 00...0 \rangle + \sin \alpha | 11...1 \rangle$. 
Consider the odd number of particles.
The non vanishing components of the correlation 
tensor of these states are:
\begin{itemize}
\item $T_{3...3} = \cos{2\alpha}$,
\item $T_{1...1} = \sin{2\alpha}$,
\item components with $2k$ indices equal to $2$ and the rest to $1$ 
(e.g., for $N=3$, $T_{122}$, etc., for $N=5$, $T_{11122}, T_{12222}$, etc.)
have the value of $(-1)^k \sin{2\alpha}$.
\end{itemize}
Take tensors $T$ and $T'$ in (\ref{N_necessary}) in the same bases. In such a case one has
\begin{eqnarray}
&&\sum_{i_1,....i_{N-1}=1,2} T_{i_1...i_{N-1}2}^2 + 
\sum_{k_1...k_{N-1}=1,3} T_{k_1'...k_{N-1}'3}^{2}  \\
&&=\cos^2{2\alpha} + \Big[\sum_{i=1}^{(N-1)/2} {N-1 \choose 2i -1} \Big] 
\sin^2{2\alpha} \\
&&=\cos^2{2 \alpha} + 2^{N-2} \sin^2{\alpha} = 1+ (2^{N-2} -1) \sin^2{2 \alpha} > 1.
\end{eqnarray}

The correlation tensor for even particle generalized GHZ states differs from the one above only in the component 
$T_{3...3}$ which is always equal to $1$.  All such states violate the new inequalities.

\section{More examples}

The following examples illustrate the usefulness of the new inequalities.

\subsection{$|W\rangle$ states}

For the $|W\rangle$ states the new inequalities are stronger than the standard
inequalities. For example three particle $|W\rangle$ state violates the standard
inequalities by the factor $1.5229$ whereas for the new inequalities the factor
is equal to $1.5275$. 

Generally N qubit $|W \rangle$ state's non vanishing correlation
tensor components are: $T_{z...z}=1$; $T_{xxz...z}=-2/N$ ( $N \choose 2$
elements, therein: ${N-2} \choose 2$ elements $T_{xxz...\textbf{z}}$ and $N-1$
elements $T_{xz...z\textbf{x}}$); $T_{yyz...z}=-2/N$ ( $N \choose 2$
elements, therein: ${N-2} \choose 2$ elements $T_{yyz...\textbf{z}}$ and $N-1$ elements
$T_{yz...z\textbf{y}}$).
The maximal value of the left hand side of (\ref{N_necessary}) is given by
\begin{eqnarray}
&&\sum_{i_1...i_{N-1}=\{x,z\}} T_{i_1...i_{N-1}\textbf{z}}^2 +
\sum_{i_1...i_{N-1}=\{y,z\}} T_{i_1...i_{N-1}\textbf{y}}^2\\
&& = 1 + {{N-1} \choose 2} (2/N)^2 + (N-1) (2/N)^2 = 3-2/N
\end{eqnarray}
Therefore the factor by which the $4\times ...\times 4\times 2$ inequalities are
violated is $\sqrt{3 - 2/N}$.

Thus if one considers noise admixture to the $|W\rangle$ states, 
in such a form that one has a mixed state $\rho_{|W\rangle} = (1-V) \rho_{noise} + V
 |W \rangle \langle W|$, with $\rho_{noise} = \hat 1 / 2^N$, then the new inequalities
show that for $V \geq 1/\sqrt{3- 2/N}$ there is no local realistic description for the
correlation functions.

This should be compared with the identical threshold for the standard inequalities \cite{ADITI},
which is only sufficient for them to hold. However, it is not equal to the necessary one. 
This clearly  illustrates the advantage of the new inequalities.

\subsection{A four photon state}

The next interesting example is the state reads \cite{WZuk}:
\begin{equation}
|\Psi \rangle = \sqrt{1/3} \Big( |0000 \rangle + |1111 \rangle +
\frac{1}{2}(|1010\rangle + |0101 \rangle + |0110\rangle + |1001\rangle) \Big),
\end{equation}
Its non vanishing correlation tensor components are:
\begin{eqnarray}
&&T_{xxxx}=T_{yyyy}=T_{zzzz}=1, \\
&&T_{xxyy}=T_{xxzz}=T_{yyxx}=T_{yyzz}=T_{zzxx}=T_{zzyy}=-\frac{1}{3}, \\
&&T_{xzxz}=T_{xzzx}=T_{zxxz}=T_{zxzx}=\frac{2}{3}, \\
&&T_{xyxy}=T_{xyyx}=T_{yxxy}=T_{yxyx}=T_{yzyz}=T_{yzzy}=T_{zyyz}=T_{zyzy}=-\frac{2}{3}.
\end{eqnarray}
Since $\sqrt{\sum_{ijk=\{x,z\}} T_{ijk\textbf{z}}^2 +
\sum_{ijk=\{y,z\}} T_{ijk\textbf{y}}^2}$ is equal to 2 the $4 \times 4 \times 4 \times 2$ 
inequalities are violated  by a factor of 2. For the standard
inequalities the maximal violation is equal to only $1.8856$.

\section{$2^{N-1} \times 2^{N-1} \times 2^{N-2} \times 2^{N-3} \times .... \times 2$ case}

The last type of inequalities presented in \cite{BLPZ} is $2^{N-1} \times 2^{N-1} \times 2^{N-2} \times 2^{N-3} 
\times ... \times 2$. The necessary condition for violation of the Bell inequality is relatively easy 
to obtain, with the use of Cauchy-Schwartz inequality. The equivalent sufficient condition for violation can 
also be obtained by using Schmidt decomposition in the way presented earlier. 
Simply the problem of maximization of the Bell expressions with rank $N$ tensor can be split into problems 
considering lower rank tensors. Eventually one gets to the problem with rank $2$ tensors which, in 
fact, was solved in \cite{Horodecki}.

W.L. and T.P. are supported 
by the UG grant BW/5400-5-0256-3 and Stypendium FNP. M.\.Z. is supported by the Subsydium Profesorskie FNP.

\end{document}